\documentclass[aps,prl,twocolumn,showpacs,superscriptaddress,groupedaddress]{revtex4}  
\usepackage{graphicx}  
\usepackage{dcolumn}   
\usepackage{bm}        
\usepackage{amssymb}   
\usepackage[tight]{subfigure}

\newcommand{\pt} {\mbox{$p_T$}}
\newcommand{\dzero}     {D0}
\newcommand{\ttbar}     {\mbox{$t\bar{t}$}}
\newcommand{\sigmattbar} {\mbox{$\sigma_{t\bar{t}}$}}

\newcommand{\pythia}    {\sc{pythia}}

\newcommand{\alpgen}    {\sc{alpgen}}

\newcommand{\geant}     {\sc{geant}}
\newcommand{\met}       {\mbox{$\not\!\!E_T$}}

\newcommand{\ljets}     {\mbox{\ensuremath{\ell {\rm +jets}}}}
\newcommand{\dilepton}  {\mbox{\ensuremath{\ell\ell}}}
\newcommand{\ejets}     {$e$+jets}
\newcommand{\mujets}    {$\mu$+jets}
\newcommand{\lumi}      {5.4~fb$^{-1}$}
\newcommand{\vtb}       {\ensuremath{V_{tb}}} 
\newcommand{\Btb}       {\mbox{\ensuremath{\mathcal{B}(t \rightarrow Wb)}}}
\newcommand{\Btq}       {\mbox{\ensuremath{\mathcal{B}(t \rightarrow Wq)}}}
\newcommand{\Rb}	{\mbox{\ensuremath{\Btb/\Btq}}}
\newcommand{\NIM}	{Nucl.\ Instrum.\ Methods  A\ }
\newcommand{\PRL}	{Phys.\ Rev.\ Lett.\ }
\newcommand{\resultrb} {0.90} 
\newcommand{\errrb}    {\pm 0.04}
\newcommand{\fullresultrb} {\ensuremath{\resultrb \errrb}}

\begin{document}

\hspace{5.2in} \mbox{FERMILAB-PUB-11-300-E}

\title{Precision measurement of the ratio {\boldmath \Rb}}
\affiliation{Universidad de Buenos Aires, Buenos Aires, Argentina}
\affiliation{LAFEX, Centro Brasileiro de Pesquisas F{\'\i}sicas, Rio de Janeiro, Brazil}
\affiliation{Universidade do Estado do Rio de Janeiro, Rio de Janeiro, Brazil}
\affiliation{Universidade Federal do ABC, Santo Andr\'e, Brazil}
\affiliation{Instituto de F\'{\i}sica Te\'orica, Universidade Estadual Paulista, S\~ao Paulo, Brazil}
\affiliation{Simon Fraser University, Vancouver, British Columbia, and York University, Toronto, Ontario, Canada}
\affiliation{University of Science and Technology of China, Hefei, People's Republic of China}
\affiliation{Universidad de los Andes, Bogot\'{a}, Colombia}
\affiliation{Charles University, Faculty of Mathematics and Physics, Center for Particle Physics, Prague, Czech Republic}
\affiliation{Czech Technical University in Prague, Prague, Czech Republic}
\affiliation{Center for Particle Physics, Institute of Physics, Academy of Sciences of the Czech Republic, Prague, Czech Republic}
\affiliation{Universidad San Francisco de Quito, Quito, Ecuador}
\affiliation{LPC, Universit\'e Blaise Pascal, CNRS/IN2P3, Clermont, France}
\affiliation{LPSC, Universit\'e Joseph Fourier Grenoble 1, CNRS/IN2P3, Institut National Polytechnique de Grenoble, Grenoble, France}
\affiliation{CPPM, Aix-Marseille Universit\'e, CNRS/IN2P3, Marseille, France}
\affiliation{LAL, Universit\'e Paris-Sud, CNRS/IN2P3, Orsay, France}
\affiliation{LPNHE, Universit\'es Paris VI and VII, CNRS/IN2P3, Paris, France}
\affiliation{CEA, Irfu, SPP, Saclay, France}
\affiliation{IPHC, Universit\'e de Strasbourg, CNRS/IN2P3, Strasbourg, France}
\affiliation{IPNL, Universit\'e Lyon 1, CNRS/IN2P3, Villeurbanne, France and Universit\'e de Lyon, Lyon, France}
\affiliation{III. Physikalisches Institut A, RWTH Aachen University, Aachen, Germany}
\affiliation{Physikalisches Institut, Universit{\"a}t Freiburg, Freiburg, Germany}
\affiliation{II. Physikalisches Institut, Georg-August-Universit{\"a}t G\"ottingen, G\"ottingen, Germany}
\affiliation{Institut f{\"u}r Physik, Universit{\"a}t Mainz, Mainz, Germany}
\affiliation{Ludwig-Maximilians-Universit{\"a}t M{\"u}nchen, M{\"u}nchen, Germany}
\affiliation{Fachbereich Physik, Bergische Universit{\"a}t Wuppertal, Wuppertal, Germany}
\affiliation{Panjab University, Chandigarh, India}
\affiliation{Delhi University, Delhi, India}
\affiliation{Tata Institute of Fundamental Research, Mumbai, India}
\affiliation{University College Dublin, Dublin, Ireland}
\affiliation{Korea Detector Laboratory, Korea University, Seoul, Korea}
\affiliation{CINVESTAV, Mexico City, Mexico}
\affiliation{Nikhef, Science Park, Amsterdam, the Netherlands}
\affiliation{Radboud University Nijmegen, Nijmegen, the Netherlands and Nikhef, Science Park, Amsterdam, the Netherlands}
\affiliation{Joint Institute for Nuclear Research, Dubna, Russia}
\affiliation{Institute for Theoretical and Experimental Physics, Moscow, Russia}
\affiliation{Moscow State University, Moscow, Russia}
\affiliation{Institute for High Energy Physics, Protvino, Russia}
\affiliation{Petersburg Nuclear Physics Institute, St. Petersburg, Russia}
\affiliation{Instituci\'{o} Catalana de Recerca i Estudis Avan\c{c}ats (ICREA) and Institut de F\'{i}sica d'Altes Energies (IFAE), Barcelona, Spain}
\affiliation{Stockholm University, Stockholm and Uppsala University, Uppsala, Sweden}
\affiliation{Lancaster University, Lancaster LA1 4YB, United Kingdom}
\affiliation{Imperial College London, London SW7 2AZ, United Kingdom}
\affiliation{The University of Manchester, Manchester M13 9PL, United Kingdom}
\affiliation{University of Arizona, Tucson, Arizona 85721, USA}
\affiliation{University of California Riverside, Riverside, California 92521, USA}
\affiliation{Florida State University, Tallahassee, Florida 32306, USA}
\affiliation{Fermi National Accelerator Laboratory, Batavia, Illinois 60510, USA}
\affiliation{University of Illinois at Chicago, Chicago, Illinois 60607, USA}
\affiliation{Northern Illinois University, DeKalb, Illinois 60115, USA}
\affiliation{Northwestern University, Evanston, Illinois 60208, USA}
\affiliation{Indiana University, Bloomington, Indiana 47405, USA}
\affiliation{Purdue University Calumet, Hammond, Indiana 46323, USA}
\affiliation{University of Notre Dame, Notre Dame, Indiana 46556, USA}
\affiliation{Iowa State University, Ames, Iowa 50011, USA}
\affiliation{University of Kansas, Lawrence, Kansas 66045, USA}
\affiliation{Kansas State University, Manhattan, Kansas 66506, USA}
\affiliation{Louisiana Tech University, Ruston, Louisiana 71272, USA}
\affiliation{Boston University, Boston, Massachusetts 02215, USA}
\affiliation{Northeastern University, Boston, Massachusetts 02115, USA}
\affiliation{University of Michigan, Ann Arbor, Michigan 48109, USA}
\affiliation{Michigan State University, East Lansing, Michigan 48824, USA}
\affiliation{University of Mississippi, University, Mississippi 38677, USA}
\affiliation{University of Nebraska, Lincoln, Nebraska 68588, USA}
\affiliation{Rutgers University, Piscataway, New Jersey 08855, USA}
\affiliation{Princeton University, Princeton, New Jersey 08544, USA}
\affiliation{State University of New York, Buffalo, New York 14260, USA}
\affiliation{Columbia University, New York, New York 10027, USA}
\affiliation{University of Rochester, Rochester, New York 14627, USA}
\affiliation{State University of New York, Stony Brook, New York 11794, USA}
\affiliation{Brookhaven National Laboratory, Upton, New York 11973, USA}
\affiliation{Langston University, Langston, Oklahoma 73050, USA}
\affiliation{University of Oklahoma, Norman, Oklahoma 73019, USA}
\affiliation{Oklahoma State University, Stillwater, Oklahoma 74078, USA}
\affiliation{Brown University, Providence, Rhode Island 02912, USA}
\affiliation{University of Texas, Arlington, Texas 76019, USA}
\affiliation{Southern Methodist University, Dallas, Texas 75275, USA}
\affiliation{Rice University, Houston, Texas 77005, USA}
\affiliation{University of Virginia, Charlottesville, Virginia 22901, USA}
\affiliation{University of Washington, Seattle, Washington 98195, USA}
\author{V.M.~Abazov} \affiliation{Joint Institute for Nuclear Research, Dubna, Russia}
\author{B.~Abbott} \affiliation{University of Oklahoma, Norman, Oklahoma 73019, USA}
\author{B.S.~Acharya} \affiliation{Tata Institute of Fundamental Research, Mumbai, India}
\author{M.~Adams} \affiliation{University of Illinois at Chicago, Chicago, Illinois 60607, USA}
\author{T.~Adams} \affiliation{Florida State University, Tallahassee, Florida 32306, USA}
\author{G.D.~Alexeev} \affiliation{Joint Institute for Nuclear Research, Dubna, Russia}
\author{G.~Alkhazov} \affiliation{Petersburg Nuclear Physics Institute, St. Petersburg, Russia}
\author{A.~Alton$^{a}$} \affiliation{University of Michigan, Ann Arbor, Michigan 48109, USA}
\author{G.~Alverson} \affiliation{Northeastern University, Boston, Massachusetts 02115, USA}
\author{G.A.~Alves} \affiliation{LAFEX, Centro Brasileiro de Pesquisas F{\'\i}sicas, Rio de Janeiro, Brazil}
% ...
%
%
\collaboration{The D0 Collaboration\footnote{with visitors from
$^{a}$Augustana College, Sioux Falls, SD, USA,
$^{b}$The University of Liverpool, Liverpool, UK,
$^{c}$SLAC, Menlo Park, CA, USA,
$^{d}$University College London, London, UK,
$^{e}$Centro de Investigacion en Computacion - IPN, Mexico City, Mexico,
$^{f}$ECFM, Universidad Autonoma de Sinaloa, Culiac\'an, Mexico,
and 
$^{g}$Universit{\"a}t Bern, Bern, Switzerland.
$^{\ddag}$Deceased.
}} \noaffiliation
\vskip 0.25cm
\date{June 27, 2011}

\begin{abstract}
We present a measurement of the ratio of top quark branching fractions
{\mbox{$R =\Rb$}}, where $q$ can be a $d$, $s$ or $b$~quark, in the lepton+jets and dilepton \ttbar\ final states.
The measurement uses data from \lumi\ of {\mbox{$p\bar p$}} collisions collected  with the \dzero\ detector at the Fermilab Tevatron Collider.
We measure $R= \fullresultrb$, and we extract the CKM matrix element $|\vtb|$ as $|\vtb| = 0.95 \pm 0.02$, assuming unitarity of the $3\times3$ CKM matrix. 
\end{abstract}

\maketitle

The standard model (SM) of particle physics contains three generations of quarks. 
The top quark belongs to the third generation, and is of interest not only because of its 
large mass~\cite{worldaveragemass}, but also because its decay has not been examined in great detail, and may prove to be inconsistent with the SM.
The decay rate of the top quark into a $W$ boson and a down-type quark $q$ {\mbox{($q$ = $d, s, b$)}} is 
proportional to $|V_{tq}|^2$, the squared element of the Cabibbo Kobayashi Maskawa (CKM)  matrix~\cite{CKM}.
Under the assumption of a unitary $3\times 3$ CKM matrix, $|\vtb|$ is highly constrained to $|\vtb|=0.999152^{ +0.000030}_{ -0.000045}$~\cite{pdg}, and
the top quark decays almost exclusively to $Wb$.
The existence of a fourth generation of quarks would remove this constraint and accomodate significantly smaller values of $|\vtb|$.
A smaller value of  $|\vtb|$  could be observed directly through the electroweak production of single top quarks, for which the cross section is proportional 
to $|\vtb|^2$, and could also affect the decay rates in the $t\bar{t}$ production channel. 
The latter can be used to extract the ratio of branching fractions $R$:
\begin{eqnarray}
\label{eq:Rdef}
R = \frac{\Btb}{\Btq} & = & \frac{\mid \vtb \mid^2}{\mid \vtb \mid^2 + \mid V_{ts}\mid^2 + \mid V_{td}\mid^2}  \;.
\end{eqnarray}
Given the constraints on the unitary $3\times 3$ CKM matrix elements, $R$ is expected to be $0.99830 ^{+0.00006}_{-0.00009}$.
Along with a measurement of $|V_{tb}|$ using single top quark production, the measurement of $R$ provides the possibility of a study of $|V_{tq}|$~\cite{Vtb_theory}.

This Letter presents a measurement of $R$ using a data sample corresponding to an integrated luminosity of \lumi\ of {\mbox{$p\bar p$}} collisions, collected 
with the \dzero\ detector at the Fermilab Tevatron $p\bar{p}$ Collider at {\mbox{$\sqrt{s}$ =\ 1.96\ TeV}}. 
We present measurements in the lepton+jets (\ljets) channel, in which one $W$ boson from 
{\mbox{$ t\overline{t} \rightarrow W^{+} q W^{-} \overline{q}$}} production decays into a quark and an antiquark and the other into a charged lepton and a neutrino, and 
in dilepton (\dilepton) final states, in which both $W$ bosons decay into $\ell \nu$.
We also present the combination of these two measurements. 
We consider events in which the charged leptons are either electrons or muons, produced directly from the $W$ decay or 
from the leptonic decay of a $\tau$ lepton. The result from the \ljets\ channel corresponds to an improvement of the measurement using $0.9~\text{fb}^{-1}$~\cite{ljetp17},
in which we extracted $R > 0.79$ at a 95\% CL. 
This is the first \dzero\  measurement of $R$  in the \dilepton\ channel.
The CDF Collaboration has measured $R$ in the \ljets\ and \dilepton\ channels in $160~\text{pb}^{-1}$ of integrated luminosity~\cite{CDFref}, and found a limit of $R > 0.61$ at 95\% CL.

Our measurement is performed by distinguishing between the standard decay mode of the top quark
$t\bar{t} \rightarrow W^{+}bW^{-}\bar{b}$ (indicated by $bb$),
and decay modes that include light quarks ($q_l = d, s$):
$t\bar{t}\rightarrow W^{+}bW^{-}\bar{q}_l$ ($bq_l$) and
$t\bar{t} \rightarrow W^{+}q_lW^{-}\bar{q}_l$ ($q_lq_l$).
The selection of an enriched \ttbar\ sample and identification of jets from $b$~quarks are crucial elements of the analysis.

The \dzero\ detector~\cite{run2det} has a central tracking system consisting of a silicon microstrip tracker and a fiber tracker,
both located within a 1.9~T superconducting solenoidal magnet, designed to optimize tracking at pseudorapidities $|\eta|<3$~\footnote{The pseudorapidity is
defined as $\eta = - \ln [\tan(\theta/2)]$, where $\theta$ is the polar angle with respect to the proton beam.}.
The liquid-argon/uranium calorimeter has a central section covering pseudorapidities $|\eta|$ up to $\approx 1.1$, 
and two end calorimeters that extend coverage to $|\eta|\approx 4.2$~\cite{run1det}.
The outer muon system, covering $|\eta|<2$, consists of a layer of tracking detectors and scintillation trigger 
counters in front of 1.8~T toroids, followed by two similar layers behind the toroids~\cite{run2muon}.
In the \ljets\ channel, we rely on the event selections used for the measurement of the \ttbar\ production cross section~\cite{xsecljet}. 
Details on object identification and selections are only briefly summarized as follows. We select $t\bar{t}$ 
events by taking advantage of their distinct topology. We require at least three jets within $|\eta|<2.5$, with 
transverse momentum $\pt>20$~GeV, of which at least one has to have $\pt>40$~GeV.
We require one electron (muon) of $\pt>20$~GeV, $|\eta|<1.1$ ($|\eta|<2.0$) isolated from jets. 
In addition, events with a second isolated electron or muon of $\pt>15$~GeV are removed in order 
to ensure that the \ljets\ and \dilepton\ samples are statistically independent. 
The imbalance in transverse energy, $\met$, must fulfill $\met>20$~GeV ($\met>25$~GeV) in the \ejets\ (\mujets) channel. 
The most important background in the \ljets\ channel is from $W$+jets events which can produce a similar final state to $t\bar{t}$ events.
There is also significant background contribution from multijet production, in which a jet is misidentified as an electron, 
or a muon from the semileptonic decay of a hadron appears isolated. 
Smaller background contributions arise from electroweak single top quark production, 
Drell-Yan and $Z$ boson production (decaying to {\mbox{$l^+l^-$+jets}})
or diboson production ($WW$, $WZ$ or $ZZ$).  
The multijet background is estimated from control samples in data~\cite{xsecljet}, 
while the $t\bar{t}$  signal and electroweak backgrounds are simulated using Monte Carlo (MC) event generators {\alpgen}~ and {\pythia}~\cite{Alpgen,Pythia}, 
and a {\geant}-based~\cite{geant} simulation of the {\dzero} detector. 
Drell-Yan and $Z$ boson production is normalized to the next-to-next-to-leading order (NNLO) QCD prediction~\cite{FEWZ}.
All other electroweak backgrounds are normalized to their next-to-leading order (NLO) cross sections,
while the $W$+jets background is normalized to data using an iterative procedure~\cite{xsecljet}.

For the \dilepton\ channel, we use the same selections as used for the measurement of the \ttbar\ cross section described in Ref.~\cite{xsecdilep}. 
In this final state, the \ttbar\ signature consists of two energetic, 
oppositely charged isolated leptons, large $\met$ and two high $\pt$ jets. 
We consider separately the three final states $ee$, $\mu\mu$ and $e\mu$. 
For the $e\mu$ final state, we also consider events with only
one reconstructed jet.
To select \ttbar\ events, we require  two isolated leptons, electrons or muons, with $\pt > 15$~GeV, $|\eta|< 1.1$ or $1.5<|\eta|<2.5$  ($|\eta|<2$) 
for the electron (muon), and at least two jets with $\pt>20$~GeV and $|\eta|<2.5$. For $\mu\mu$ events we require $\met>40$~GeV. 
In the $e\mu$ channel, the sum of the transverse momenta of the lepton and two jets of highest $\pt$ must be 
larger than $110$~GeV. 
That sum must be higher than $105$~GeV when only one jet is reconstructed.
In the $ee$ and $\mu\mu$ channels we use the $\met$ significance to differentiate events with true $\met$ from escaping neutrinos and events 
with $\met$ arising from mismeasurement.
The  $\met$ significance for each event is defined in terms of a likelihood discriminant constructed from the ratio of $\met$ to its uncertainty~\cite{metsig}.
The significance is required to correspond to more than five.
The main background in the \dilepton\ final states is composed of Drell-Yan, $Z$ boson production and diboson events,
and is estimated using MC simulation, normalized to the NNLO and NLO cross sections respectively.
There is also a background from multijet events that we estimate from data~\cite{xsecdilep}. 

We use a neural network (NN) $b$-tagging algorithm~\cite{bidNIM} to identify jets that contain $b$ quarks, 
and thereby distinguish the $bb$, $bq_l$ and $q_lq_l$ $t\bar{t}$ final states.
The inputs to the NN include impact parameters of tracks associated with the jets,
and the properties of secondary vertices within the jet.
Only taggable jets,  i.e., jets matched to a set of tracks, are considered by the NN. 
For each taggable jet, we obtain an output from the NN which ranges between zero and twelve,
with larger values more likely to correspond to jets originating from $b$ quarks.
Non-taggable jets are assigned the NN output value $-1$. 

\begin{figure*}[ht]
\begin{center}
\setlength{\unitlength}{1.0cm}
\begin{picture}(18.0,5.2)
\put(0.1,0.2){\includegraphics[width=5.7cm]{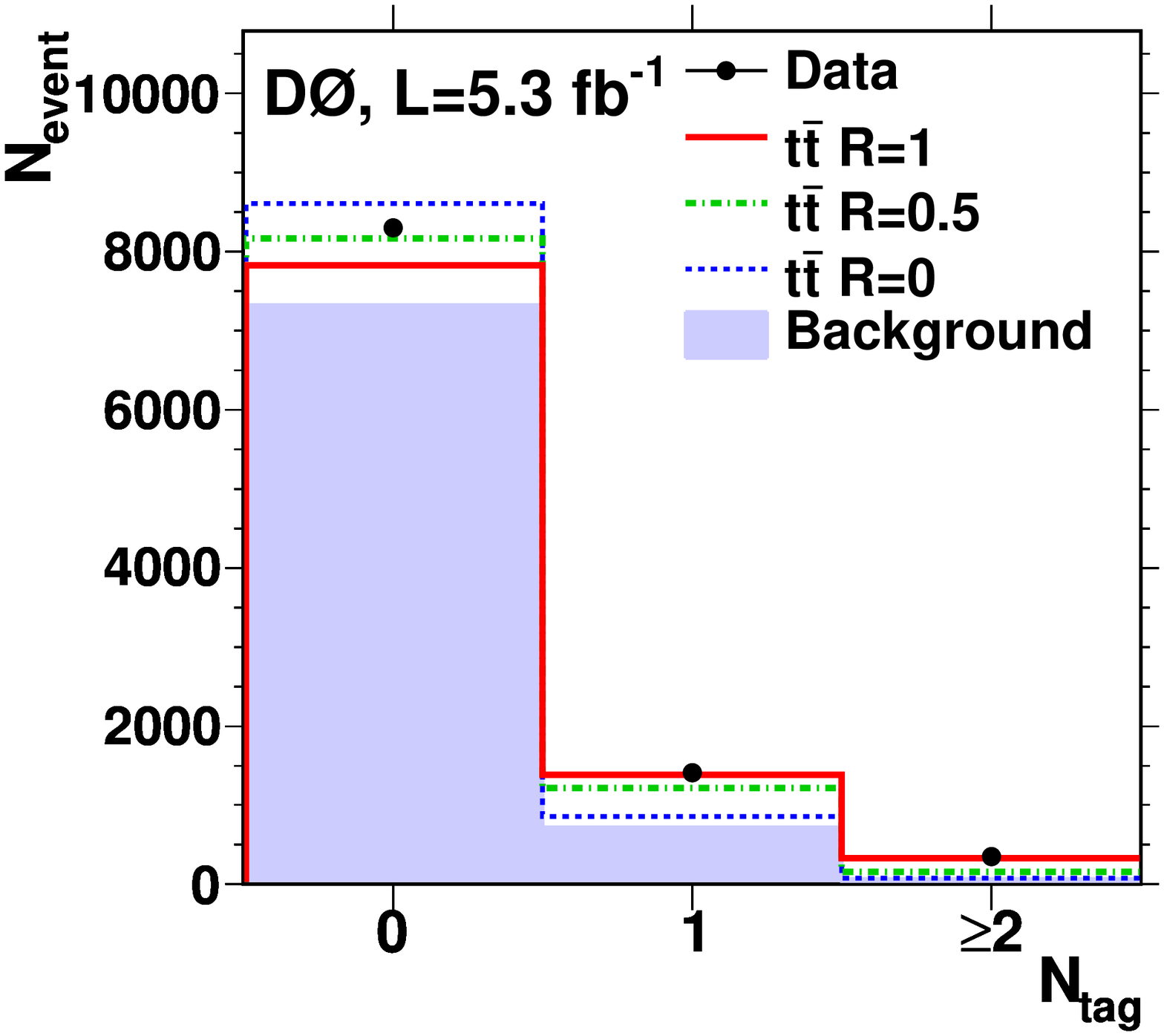}  }
\put(6.1,0.2){\includegraphics[width=5.7cm]{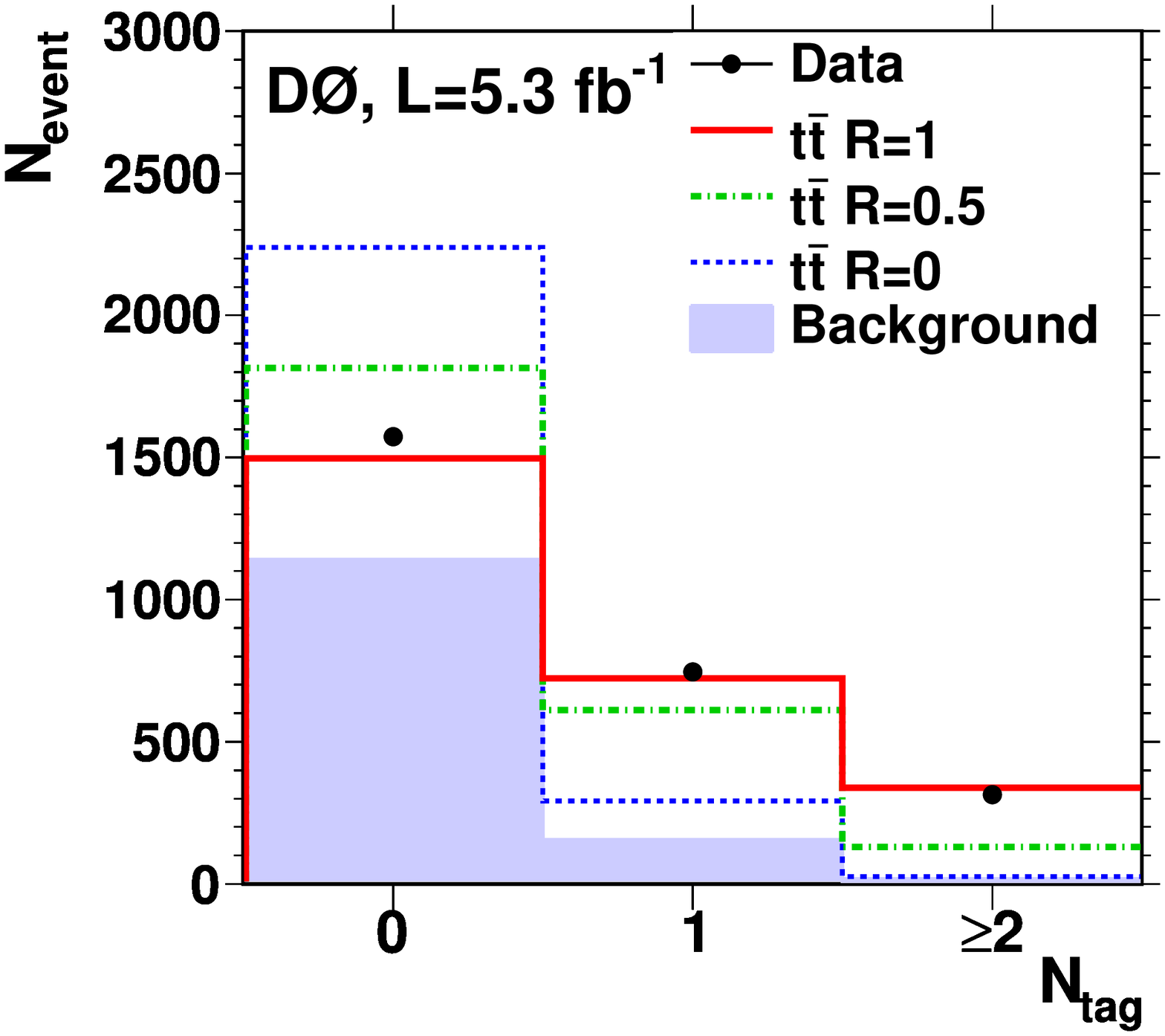} }
\put(12.1,0.2){\includegraphics[width=5.45cm]{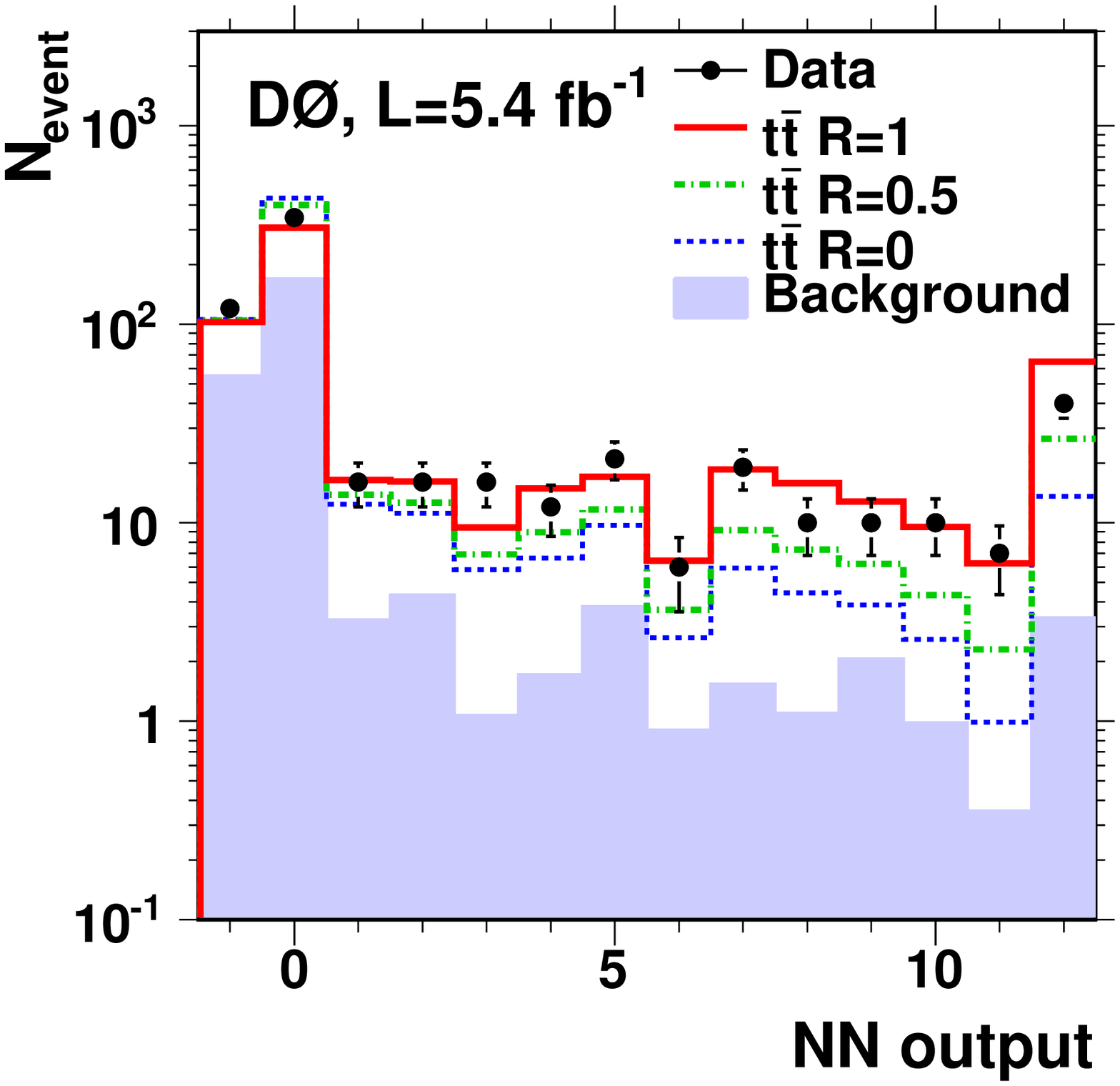} }
\put(4.9, 4.6){\bf(a)}
\put(10.9, 4.6){\bf(b)}
\put(16.7, 4.6){\bf(c)}
\end{picture}
\caption{\label{fig:njets}{(color online)} (a) Number of $b$-tagged jets in \ljets\ events with three jets and
(b) at least four jets. (c) Distribution in the minimum $b$-tag NN output of the jets of highest-$\pt$ for dilepton final states.}
\end{center}
\end{figure*}

We pursue different strategies to measure $R$ in the \ljets\ and \dilepton\  channels.
In the \ljets\ channel, we count the number of jets that pass our threshold on the $b$-tagging NN output;
this requirement has an efficiency for $b$ jets of 55 $\pm$ 4\%, while admitting 1.5 $\pm$ 0.1\% of light jets.
The events are divided into subsamples according to lepton flavor ($e$ or $\mu$), the number
of jets in the event ($3$ and $>3$), the data taking period (the first $1$~fb$^{-1}$ and the rest~\cite{xsecljet}), and the number
of identified $b$ jets (0, 1 or $> 1$).
The events are separated further using a multivariate kinematic discriminant in subsamples dominated by background, 
i.e., events with zero $b$-tagged jets, or one $b$-tagged jet in the sample with  exactly three jets, and zero $b$-tagged jets in the sample with more than three jets.
This discriminant is based on a multivariate technique
(random forest of decision trees~\cite{TMVA}) and uses several variables that exploit the
kinematic differences between \ttbar\ signal and background. 
In addition to $t\bar{t}$ MC samples with SM decay $t\bar{t} \to WbWb$,
samples for the decay modes including light quarks are generated with {\pythia} 
($t\bar{t} \to WbWq_l$ and $t\bar{t} \to Wq_lWq_l$), for a top quark mass of $m_{t}=172.5$~GeV.
The expected number of \ttbar\ events with $m$ $b$-tagged jets can be written as:
\begin{eqnarray}
\mu_{\ttbar}^m(R,\sigma_{t\bar{t}}) = [R^2\epsilon^m(bb) + 2R(1-R)\epsilon^m(bq_l) \nonumber \\
 \mbox{}+ (1-R)^2\epsilon^m(q_lq_l)]\ \sigma_{\ttbar} \mathcal{B}^2(t \to Wq) L\ ,
\label{eq:predictionR}
\end{eqnarray}
where $\epsilon^m$ is the product of the selection efficiency and the probability of an event to have $m$ $b$-tagged jets for each of the three
($bb$, $bq_l$ and $q_lq_l$) decay modes, \sigmattbar\ is the \ttbar\ production cross section and $L$ is the integrated luminosity.
A maximum likelihood fit is performed using the function:
\begin{eqnarray}
\mathcal{L}_{\ljets} = \prod_{i=1}^{N_{ch}} P[n^i,\mu^i(R,\sigma_{t\bar{t}}, \nu_k)] P[n_{MJ}^i,\mu_{MJ}^i] \times \nonumber \\
\prod_{k}\mathcal{G}(\nu_k; 0, {\rm SD})\,
\label{eq:likelihood_ljet}
\end{eqnarray}
where $i$ runs over the subsamples and bins of the multivariate discriminant, and $P[n, \mu(R,\sigma_{t\bar{t}},\nu_k)]$ is the
Poisson probability to observe $n$ events for an expected number of  $\mu(R,\sigma_{t\bar{t}},\nu_k)$ events. The expectation $\mu(R,\sigma_{t\bar{t}},\nu_k)$ is the sum of the
expected number of $t\bar{t} \to bb$, $bq_l$ and $q_lq_l$ events and the expected number of background events.
The observed and expected numbers of multijet events are denoted $n_{MJ}^i$ and $\mu_{MJ}^i$, 
and the Poisson terms $P[n_{MJ}^i,\mu_{MJ}^i]$ take into account 
the fluctuation of the number of multijet events within the statistical uncertainties with which it is determined in dedicated data samples.
Figures~\ref{fig:njets}~(a) and~(b) show the number of $b$-tagged jets in \ljets\ events for data and simulation for $R=0$, $R=0.5$ and $R=1$.
To reduce the dependence of the measurement on the input \ttbar\ cross section, we simultaneously extract \sigmattbar\ from data, 
taking into account the three channels $t\bar{t} \to bb$, $bq_l$ and $q_lq_l$.
A parameter $\nu_k$ that accounts for each independent source of systematic uncertainty $k$ is modeled by a Gaussian function $\mathcal{G}$
with a mean of zero and a width corresponding to one estimated standard deviation (SD) of that uncertainty.
This procedure correlates systematic uncertainties among channels 
by using the same parameter for a common source of systematic uncertainty.

In the dilepton channels $ee$, $\mu\mu$, and $e\mu$ with at least 2 jets, we apply the NN $b$-tagging algorithm to the two jets of highest-$p_T$, 
and use the smaller
    of the two NN outputs to calculate the likelihood. 
as it yields the best expected precision on $R$ for values close to unity.
The $b$-tagging algorithm is applied to the single reconstructed jet in the $e\mu$ channel with exactly 1 jet.
We construct the templates for the decay modes $bb$, $bq_l$, $q_lq_l$ for $t\bar{t}$ as well as for all background components, 
forming the likelihood by running the product of Eq.~\ref{eq:likelihood_ll} over all fourteen bins of the NN discriminant in all four channels, 
yielding thereby a product with 56 factors:
\begin{equation}
\mathcal{L}_{\dilepton} = \prod_{i=1}^{N_{ch}} P[n^i,\mu^i(R,\sigma_{t\bar{t}},\nu_k)] \prod_{k}\mathcal{G}(\nu_k; 0, {\rm SD}) \;.
\label{eq:likelihood_ll}
\end{equation}
The expected number of events, $\mu_{t\bar{t}}^m(R,\sigma_{t\bar{t}},\nu_k)$, is given
by Eq.~\ref{eq:predictionR}, where $\epsilon^m$ describes now the efficiency for the discriminant bin $m$, and $\nu_k$ can affect the individual components of $\mu_{t\bar{t}}^m(R,\sigma_{t\bar{t}})$.
Figure~\ref{fig:njets}~(c) compares the distributions of the discriminant for predicted and observed events in the combined \dilepton\  final state.

Several systematic uncertainties can impact the measurement of $R$.  
We consider the same sources of systematic uncertainties as for the cross-section measurements in the \ljets\ and \dilepton\ channels, and 
refer to Refs.~\cite{xsecljet,xsecdilep} for details.
The main source of systematic uncertainty on $R$ is from the  $b$-tagging probability.
Other important contributions to the systematic uncertainty on $R$ arise from the jet identification efficiency, jet energy scale and resolution, 
and uncertainties on the background normalization as well as on modeling of the signal. 
The latter includes contributions from higher order effects, color reconnection, choice of parton distribution functions and initial and final-state gluon radiation. 
For consistency with Refs.~\cite{xsecljet,xsecdilep} we also quote separately the smaller systematic contributions from limited number of events in the templates
and the uncertainties on the heavy-flavor fraction for the $W$+jets process, the trigger efficiency and lepton identification.
We account for the fact that uncertainties from jet identification, jet energy scale and resolution, $b$-jet identification, 
and higher-order corrections can affect the distribution of the discriminants in the \ljets\ channel, 
and the NN discriminants in the \dilepton\ channel.
We verify that the measurement of $R$ does not depend on $m_t$ by generating MC samples at different $m_t$ values.
In the \ljets\ channel we obtain: \\
\centerline{{\mbox{$R = 0.95 \pm 0.07 ~\text{(stat+syst)}$}} }
\centerline{{\mbox{$\sigma_{t\bar{t}} = 7.90^{+0.79}_{-0.67}~\text{(stat+syst)}$ pb}},}
and in the \dilepton\ channel \\
\centerline{{\mbox{$R = 0.86 \pm 0.05~\text{(stat+syst)}$}}  }
\centerline{{\mbox{$\sigma_{t\bar{t}} = 8.19^{+1.06}_{-0.92} ~\text{(stat+syst)}$ pb}}.}
The results are in agreement with each other, and the extracted cross sections are consistent with those from Refs.~\cite{xsecljet,xsecdilep}.
In these \sigmattbar\ measurements, we do not assume that $\Btb=1$ as was done for the results in Refs.~\cite{xsecljet,xsecdilep},
but only require $\Btq=1$.
The combined measurement is obtained by fitting simultaneously all channels in the \dilepton\ and \ljets\ final states. This yields: \\
\centerline{{\mbox{$R = \fullresultrb~\text{(stat+syst)}$}}  }
\centerline{{\mbox{$\sigma_{t\bar{t}} = 7.74^{+0.67}_{-0.57}~\text{(stat+syst)}$ pb}}.}
Table~\ref{tab:syst} summarizes the systematic uncertainties for the three results on $R$. While in the \dilepton\  channel the statistical 
uncertainty still dominates, the \ljets\ and the combined result are dominated by systematic uncertainties.  
If we assume unitarity of the CKM matrix, we extract $|\vtb| = 0.95 \pm 0.02$.
Constraining the $t\bar{t}$ cross section to the SM value of $7.5 ^{+0.6} _{-0.7}~\text{pb}$~\cite{mochuwer} yields $R=0.90 \pm 0.04$, 
identical within rounding errors to the result of the simultaneous fit.

\begin{table*}
 \begin{center}
  \caption{Uncertainties on the measurements of $R$ in the \dilepton\  and \ljets\ channels as well as for the combination of the two.
	  We evaluate the impact of each class of systematic uncertainties by calculating
	  $R$ and $\sigma_{t\bar{t}}$ using the corresponding parameters $\nu$ shifted by $\pm1$SD from their fitted mean.
          The final line shows the quadratic sum of the systematics, which can be slightly different from the one obtained with the global fit.
}
  \begin{tabular}{lcc|cc|cc} \hline
         & \multicolumn{2}{c|}{$\ell\ell$} & \multicolumn{2}{c|}{\ljets} & \multicolumn{2}{c}{Combination} \\ \hline
 Source                                         &   $+$SD  &    $-$SD   &   $+$SD   &    $-$SD  &   $+$SD  &    $-$SD  \\ \hline
 Statistical	 				&  0.041 & $-0.042$ & 0.030 & $-0.029$ & 0.023 & $-0.023$ \\	
\hline
 Muon identification 				&  0.002 & $-0.002$ & 0.000 & $-0.001$ & 0.001 & $-0.001$ \\
 Electron identification and smearing 		&  0.004 & $-0.004$ & 0.000 & $-0.000$ & 0.001 & $-0.002$ \\
 Signal modeling 				&  0.007 & $-0.006$ & 0.009 & $-0.011$ & 0.004 & $-0.006$ \\
 Triggers 					&  0.003 & $-0.003$ & 0.001 & $-0.001$ & 0.002 & $-0.002$ \\
 Jet energy scale 				&  0.008 & $-0.008$ & 0.017 & $-0.016$ & 0.003 & $-0.008$ \\
 Jet reconstruction and identification 		&  0.010 & $-0.009$ & 0.018 & $-0.022$ & 0.009 & $-0.013$ \\
 $b$-tagging 					&  0.018 & $-0.019$ & 0.065 & $-0.056$ & 0.034 & $-0.033$ \\
 Background normalization 			&  0.020 & $-0.020$ & 0.004 & $-0.005$ & 0.008 & $-0.010$ \\
 W fractions matching + higher order effects 	&     -  &      -   & 0.001 & $-0.001$ & 0.001 & $-0.002$ \\
 Instrumental background 			&  0.013 & $-0.013$ & 0.003 & $-0.004$ & 0.005 & $-0.007$ \\
 Luminosity 					&  0.010 & $-0.010$ & 0.001 & $-0.001$ & 0.004 & $-0.004$ \\
 Other 						&  0.002 & $-0.002$ & 0.000 & $-0.000$ & 0.001 & $-0.001$ \\
 Template statistics for template fits 		&  0.002 & $-0.002$ & 0.011 & $-0.011$ & 0.010 & $-0.010$ \\

\hline
 Quadratic sum of systematics 				&  0.035 & $-0.035$ & 0.071 & $-0.064$ & 0.038 & $-0.040$ \\
\hline
\end{tabular}

  \label{tab:syst}
 \end{center}
\end{table*}

\begin{figure}
\includegraphics[width=0.35\textwidth]{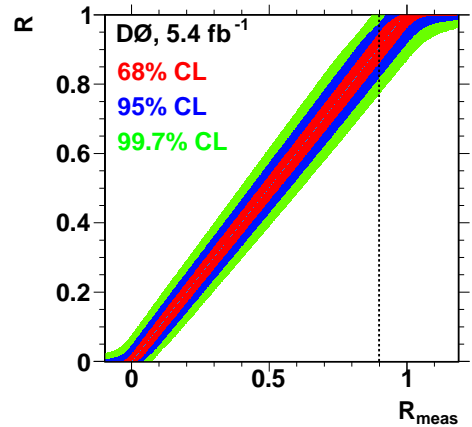}
\caption{\label{fig:FC} {(color online)} Limit bands at 68$\%$, 95$\%$ and 99.7$\%$ CL on $R$, with the measured value (dotted line).}
\end{figure}

Using the combined result, we extract intervals on $R$ as well as on $|V_{tb}|$ from Eq.~(\ref{eq:Rdef}), 
assuming unitarity of the $3\times3$ CKM matrix.
By applying the frequentist approach using the likelihood ratio ordering principle proposed by Feldman and Cousins~\cite{feldmancousins}, 
we obtain the intervals in $R$ as 0.82--0.98 and $\vtb$ as 0.90--0.99 at 95\% CL.
The expected limits are $R > 0.92$ and $\vtb > 0.96$ at 95\% CL.
Figure~\ref{fig:FC} shows the bands for 68\%, 95\% and 99.7$\%$ confidence limits on $R$.
Our result is compatible with the SM expectation at the 1.6\% level. At 99.7\% CL, we obtain
$R > 0.77$ and $|V_{tb}| > 0.88$.

Without assumptions on the unitarity of the CKM matrix, we can write Eq.~(\ref{eq:Rdef}) as: 
{\mbox{$(1-R)/R = (|V_{ts}|^2 + |V_{td}|^2)/|\vtb|^2,$}} and set a limit on this ratio at 99.7\% CL of:
{\mbox{$(1-R)/R < 0.30 $.}} \\

To summarize, we have measured the ratio of branching fractions $R = \Rb$ in both lepton+jets and dilepton channels.
In the combined analysis, we find $R=\fullresultrb$, which agrees within approximately 2.5 standard deviations with the SM
prediction of $R$ close to one. This is the most precise determination of $R$ to date. 
Using the approach of Ref.~\cite{feldmancousins} and assuming the unitarity of the CKM matrix, we extract the interval
at 95\% CL on the element \vtb\ as 0.90--0.99.

%
We thank the staffs at Fermilab and collaborating institutions,
and acknowledge support from the
DOE and NSF (USA);
CEA and CNRS/IN2P3 (France);
FASI, Rosatom and RFBR (Russia);
CNPq, FAPERJ, FAPESP and FUNDUNESP (Brazil);
DAE and DST (India);
Colciencias (Colombia);
CONACyT (Mexico);
KRF and KOSEF (Korea);
CONICET and UBACyT (Argentina);
FOM (The Netherlands);
STFC and the Royal Society (United Kingdom);
MSMT and GACR (Czech Republic);
CRC Program and NSERC (Canada);
BMBF and DFG (Germany);
SFI (Ireland);
The Swedish Research Council (Sweden);
and
CAS and CNSF (China).
%

\end{document}